\documentclass[twocolumn,aps,prl,final,superscriptaddress,floatfix,amsmath,amssymb]{revtex4}

\usepackage{graphicx}
\usepackage{dcolumn}% Align table columns on decimal point
\usepackage{bm}% bold math

\usepackage{endnotes}
\usepackage{longtable}

\usepackage{hyperref}

\newcommand{\fig}[1]{Fig.~\ref{#1}}
\renewcommand{\vec}[1]{\mathbf{#1}}

\begin{document}

\title{Excitation of coherent phonons in the one-dimensional Bi(114) surface}

\author{D.  Leuenberger}
%\affiliation{Physics Institute, University of Z\"{u}rich, Winterthurerstrasse 190, CH-8057 Z\"{u}rich, Switzerland}

\author{H.  Yanagisawa}
\altaffiliation[Present address:  ]{Institute of Quantum Electronics, Swiss Federal Institute of Technology, Wolfgang-Pauli-Strasse 16, 8093 Z\"{u}rich, Switzerland}
%\affiliation{Physics Institute, University of Z\"{u}rich, Winterthurerstrasse 190, 8057 Z\"{u}rich, Switzerland}

\author{S.  Roth}
\affiliation{Physics Institute, University of Z\"{u}rich, Winterthurerstrasse 190, 8057 Z\"{u}rich, Switzerland}

\author{J.H. Dil}
\affiliation{Physics Institute, University of Z\"{u}rich, Winterthurerstrasse 190, 8057 Z\"{u}rich, Switzerland}
\affiliation{Swiss Light Source, Paul-Scherrer Institute, 5232 Villigen PSI, Switzerland}

\author{J.W. Wells}
\affiliation{Department of Physics, Norwegian University of Science and Technology, 7491 Trondheim, Norway}

\author{P. Hofmann}
\affiliation{Department of Physics and Astronomy and Interdisciplinary Nanoscience Center (iNANO), Aarhus University,  8000 {\AA}rhus C, Denmark}

\author{J. Osterwalder}

\author{M.  Hengsberger}
\affiliation{Physics Institute, University of Z\"{u}rich, Winterthurerstrasse 190, 8057 Z\"{u}rich, Switzerland}

\date{\today}

\begin{abstract}
We present time-resolved photoemission experiments from a peculiar bismuth surface, Bi(114). The strong one-dimensional character of this surface is reflected in the Fermi surface, which consists of spin-polarized straight lines. Our results show that the depletion of the surface state and the population of the bulk conduction band after the initial optical excitation persist for very long times. The  disequilibrium within the hot electron gas along with strong electron-phonon coupling cause the excitation of phonon standing waves, which, in turn are reflected in coherent modulations of the electronic states. Beside the well-known $A_{1g}$ bulk phonon mode at 2.76~THz the time-resolved photoelectron spectra reveal a second mode at 0.72~THz which can be attributed to a standing  optical phonon mode along the atomic rows of the Bi(114) surface. 
\end{abstract}


\maketitle

Reduced  dimensionality at solid surfaces results in unique characteristics of the electronic structure as compared to the bulk. In the extreme case of a recently discovered phase of solids, the so-called topological insulators, the topology of the bulk band structure demands the existence of topologically protected metallic states at the solid surface \cite{hsieh2008,hasan2010}.  
Bismuth, a semi-metal on the verge of being a topological insulator, also features metallic surface states on all surfaces studied so far, giving a strong enhancement of the metallic density of states (DOS) at the surface  \cite{hengsberger2000JEPB, agergaard2001, ast2001, hofmann2006}.  The degeneracy of the surface states  in momentum-space is lifted  due to the strong spin-orbit coupling in bismuth and the broken inversion symmetry  at the surface; this effect is called the Rashba-effect \cite{hofmann2006}. 
In particular, the strongly anisotropic (114)-surface was found to support a spin-split quasi-one-dimensional (1D) metallic surface state \cite{wells09}. The momenta $\vec{k}_F$ of the states at the Fermi energy $E_F$ form straight lines close to the $\overline{\Gamma}$-points of the surface reciprocal lattice and  perpendicular to the direction $[1\overline{1}0]$ of the atomic rows.
In such a topology, any low-energy perturbation providing the momentum required to connect two parallel sections of the Fermi surface, generates a strong electronic response. The general response function can be calculated using the Linhard function. It becomes singular at momenta connecting long parallel sections of the Fermi surface. This leads to instabilities of the system, which may appear as charge- or spin-density waves, for instance \cite{gruener1994}. 
Previously, such a charge-density wave (CDW) ground state was proposed for the Bi(111)-surface \cite{ast2003}. On that surface the hexagonal shape of the Fermi contour of the electron pocket at $\overline{\Gamma}$ provides straight, parallel sections at opposite momenta $\pm \vec{k}_F$. A combined photoemission and tunneling microscopy study, however, provided no evidence for a CDW \cite{kim2005}. Kim and co-workers argued that due to spin-orbit interaction, the states at opposite momenta carry opposite spin, and any scattering between the states at these momenta requires spin-flip.

\begin{figure*}[!th]
\centerline{\includegraphics[width = 0.9\textwidth]{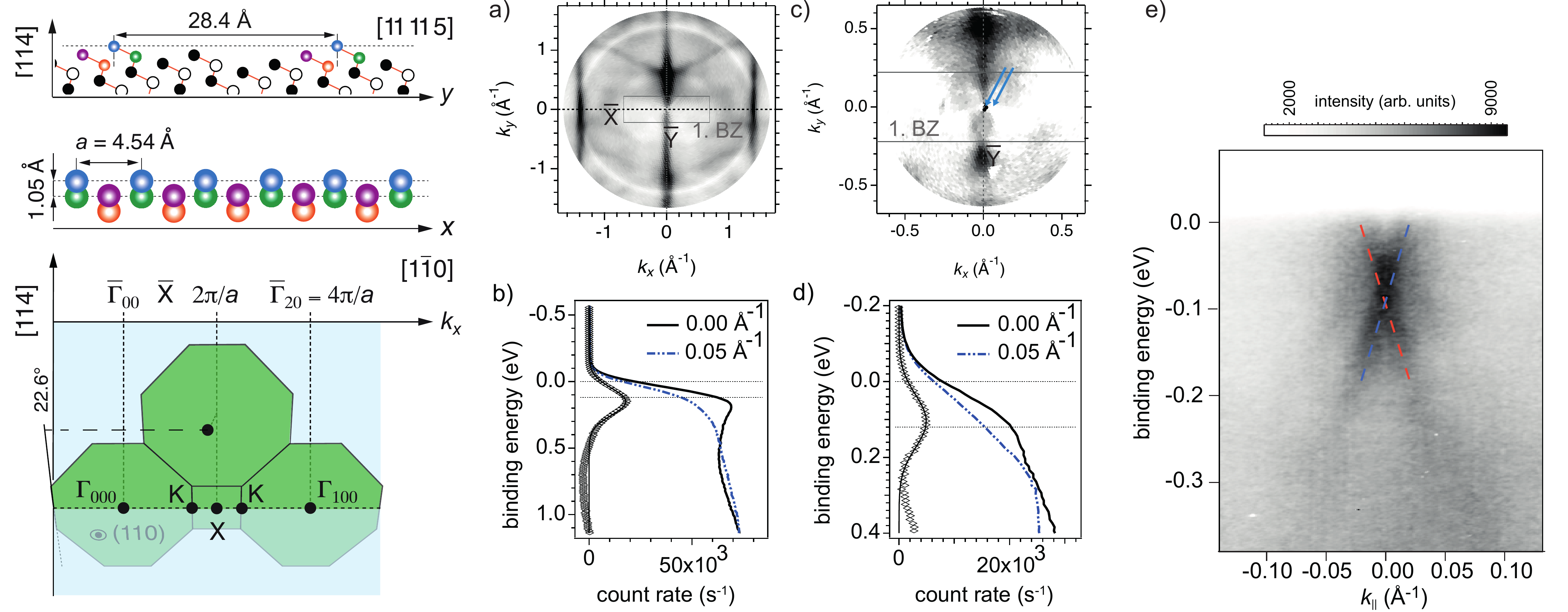}}
\caption{(Colour online) Sketch of the atomic structure of (1$\times2$)-reconstructed Bi(114); the atoms forming the rows are highlighted; bottom: Brillouin zones for surface and bulk in the plane of the binary axis and the surface normal; the projection of the bulk $X$-point along the binary axis corresponds to the center of the second surface BZ $\Gamma_{10}$. Note that the green shaded (110) plane is tilted out of the image plane. Panel (a)-(e): Photoemission data taken at the Fermi level with $h \nu$~=~21.2~eV (a and b) and 2PPE with  $2\times3.1$~eV (c and d): (a) and (c) Fermi surfaces; the surface BZ is indicated by a black rectangle. Black represents high intensity. (b) and (d)  Two selected  spectra taken for momenta indicated by the blue arrows in (c); the
black open diamonds in (b) and (d) show the difference between the two spectra, highlighting the surface state  at  $\Gamma$. Note the different energy and momentum scales in the 2PPE data. (e) Band dispersion recorded using  $h \nu$~=~21.2~eV (spin-integrated). The splitting at $E_F$ is clearly visible. The dashed lines serve as guides to the eye, the color code mimics the orthogonal spin polarisations. }
\label{FSM}
\end{figure*}

Thus, the situation on Bi(114) is particularly interesting: The Fermi surface topology renders the metallic surface states instable against charge-density wave transitions, because the complete Fermi surface of the surface states can be nested by one momentum vector.
Moreover, the distorted rhombohedral lattice structure of bulk bismuth makes it sensitive to displacive excitations  of coherent optical phonons (DECP)  \cite{cheng1990, hase1996, tinten03, fritz07, johnson2009}. Such coherent oscillations feed strong electronic modulations in space and time giving evidence for a particularly strong interplay of electron states and phonons. On the other hand, due to spin-orbit interaction, the spins of states at the momenta $+\vec{k}_F$ and $-\vec{k}_F$ have anti-parallel orientation \cite{wells09}, which inhibits scattering between these  states and, thereby, the stabilization of a standing wave. 
The Bi(114) surface therefore offers the unique possibility to study the competing effects of spin-momentum locking, reduced dimensionality,  electron-phonon coupling, and Fermi surface topology.

The Bi(114) crystal was cleaned \textit{in situ} by cycles of argon sputtering and  annealing at 300~K \cite{wells09}.  The experimental pump-probe setup was described in detail elsewhere \cite{leuenberger2011}. Briefly, we  excite the sample with a linearly $p-$polarized infrared pump pulse ($h\nu_{1}$~=~1.55~eV). The duration of the pump pulse $\Delta t_{pump}$ could be tuned by means of a grating compressor. In the measurements shown here, the absorbed pump fluence was set to $0.57$~mJ/cm$^{2}$, corresponding to an excitation density of $n \approx 0.34\%$  of the valence electron density in bulk bismuth \cite{fritz07}. The   electronic structure  was  probed by two-photon photoemission  ($2\times h\nu_{2}$~=~$2\times $3.1~eV, $\Delta t_{probe}$~=~50~fs), and the photoelectrons were detected using a hemispherical electron analyzer \cite{Greber:1997p68}. The combined energy resolution of light and spectrometer was about 50~meV, the angular resolution was set to $\pm 1^{\circ}$. All measurements were performed at room temperature.

The Bi(114) surface consists of straight atomic rows running along $[1\overline{1}0]$, which are separated by 28~\AA\  wide valleys, as shown in Fig.~\ref{FSM} \cite{wells09}. The one-dimensional spatial character  is  reflected in the surface electronic structure. The Fermi surface maps in Figs.~\ref{FSM}(a) and (c) reveal straight  lines parallel to $\overline{\Gamma}\overline{Y}$, which were attributed to a spin-split 1D surface state \cite{wells09}. All other features derive from bulk states. They disperse according to the bulk BZ and possess a defined parity and symmetry only with respect to the remaining mirror plane $yz$. 
The surface state is localized perpendicular to the atomic chains and strongly dispersing  along the chains ($\overline{\Gamma}\overline{X}$). Energy spectra  taken for various momenta along $\overline{\Gamma}\overline{X}$ reveal the surface state on top of  $\Lambda$-shaped bulk bands, and  with the band bottom at a binding energy of 100~meV,  as shown in Figs.~\ref{FSM}(b), (d), and (e). 
The agreement between the conventional one-photon photoemission and low-energy 2PPE data, shown in  Figs.~\ref{FSM}(a)-(b) and (c)-(d), respectively, ensures that the low-energy  data contain no contribution from intermediate states in the unoccupied regime between $E_{F}$ and the vacuum level \cite{hengsberger08}. 

\begin{figure*}[t]
\centerline{\includegraphics[width = 0.9\textwidth]{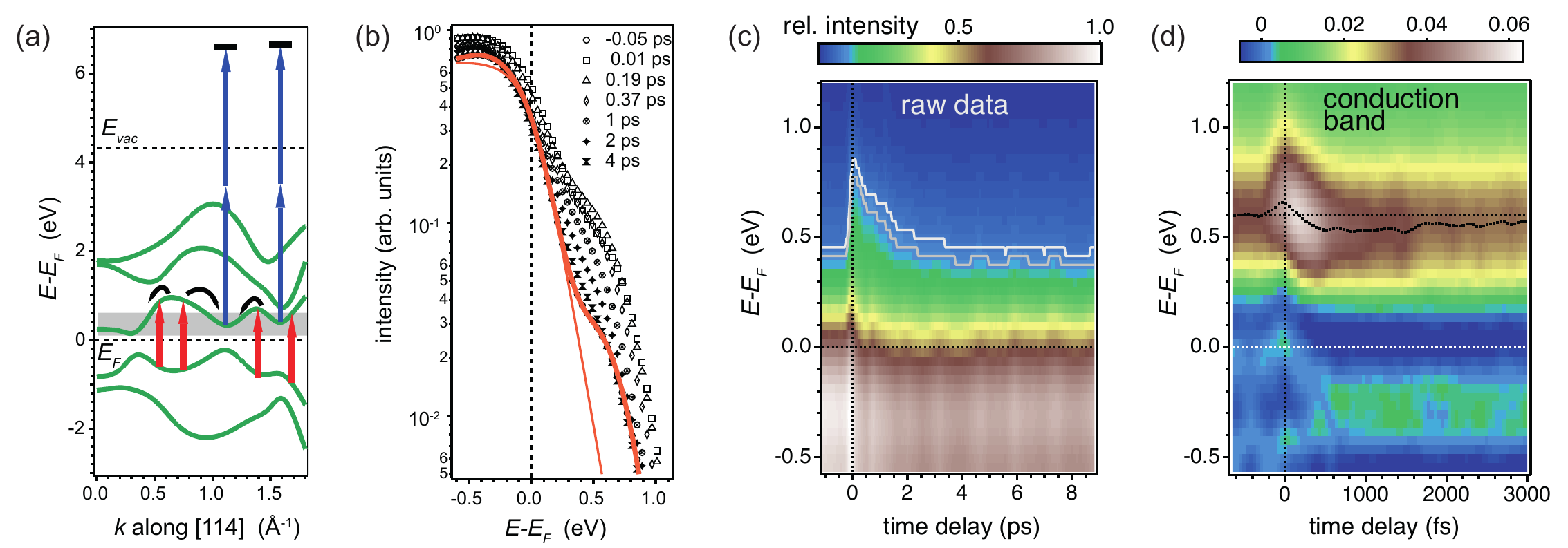}}
\caption{(Color online) Time-resolved photoemission data: (a) Momentum conserving transitions from the 5th to the 6th valence band along the [114]-direction, which can contribute to the hot electron population observed close to $\overline{\Gamma}$; the green lines denote the energy bands calculated in a tight-binding scheme \cite{liu1995}; electrons are excited by means of 1.55~eV photons (short red arrows) and accumulate in a minimum of the  conduction band  (grey shaded energy range), where they are probed by 2PPE (blue arrows). (b) Photoelectron spectra (symbols) taken close to $\overline{\Gamma}$ for various pump probe delays between -0.05~ps and 4~ps showing the pronounced signature of hot electrons above $E_{F}$; The thick (thin) orange line denotes an exemplary fit using two Gaussians multiplied with a Fermi-Dirac distribution (bare Fermi-Dirac distribution).  (c) Plot of the photoemission intensity as function of energy and time delay. (d) Same data after subtraction of the hot Fermi-Dirac distribution for each individual delay in order to highlight the transient occupation of the  conduction band. The black line follows the transient energy position of the bulk conduction band.}
\label{raw_data}
\end{figure*}

In Fig.~\ref{raw_data}, photoemission data are shown as function of time delay, taken close to normal emission ($k_{x}$~=~$k_F$~=~0.037~\AA$^{-1}$. Note that only the surface state contributes to the intensity at  the Fermi level. At zero pump-probe delay and  throughout the entire Brillouin zone, the bulk conduction band  is populated with a transient hot electron population following absorption of the infrared pump pulse. A few possible momentum-conserving optical transitions along  [114] are depicted in Fig.~\ref{raw_data}(a).
The electrons excited into the conduction band relax by scattering and accumulate at the band bottom, where they appear as a broad feature in the spectra, Fig.~\ref{raw_data}(b).\cite{footnoteCBM}  The evolution of the hot electron distribution can be seen in the false color plots in Figs.~\ref{raw_data}(c) and (d). 

\begin{figure}[!h]
\centerline{\includegraphics[width = 0.44\textwidth]{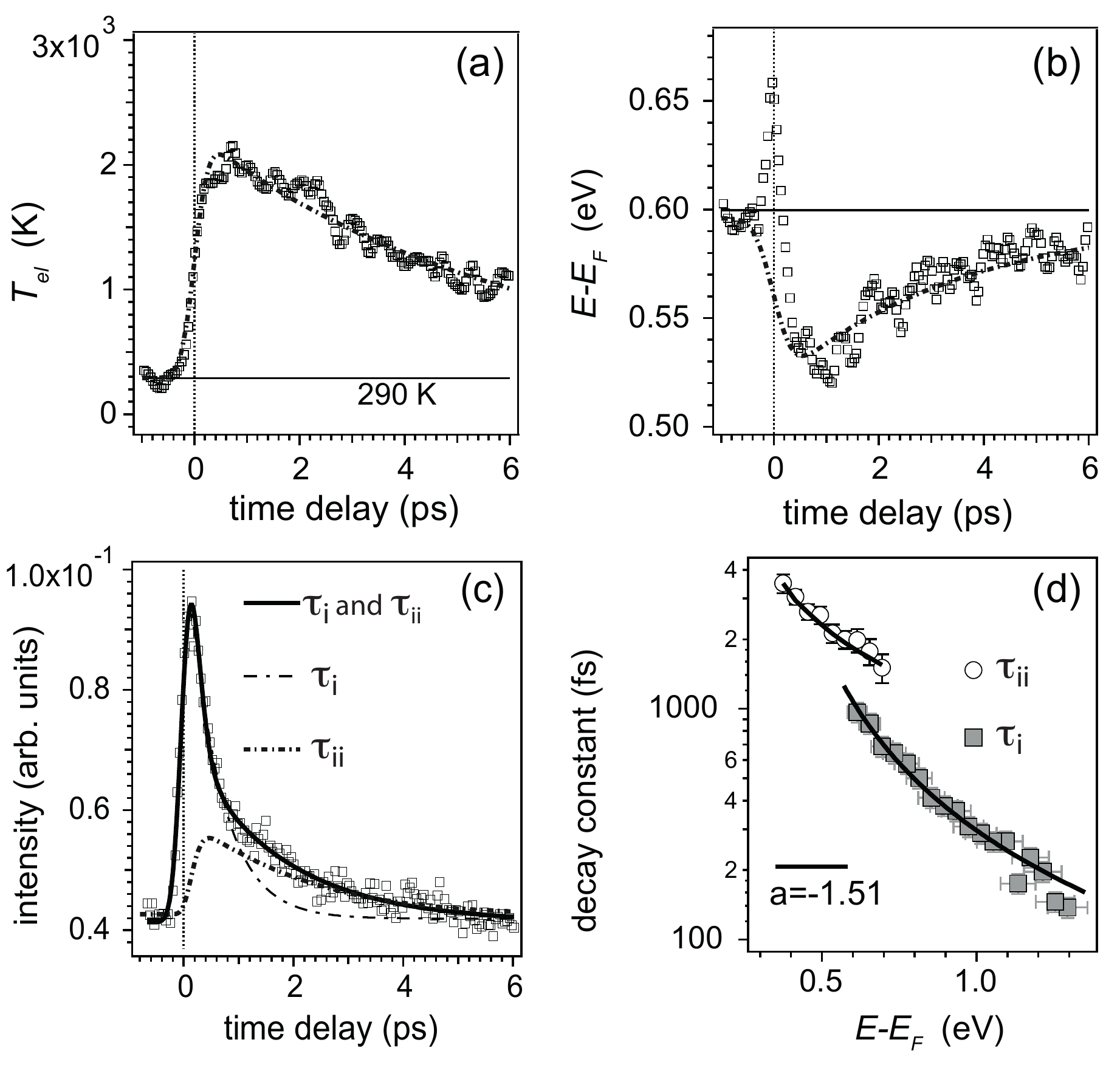}}
\caption{Analysis of hot electron dynamics: (a) Hot electron temperature $T_{el}$ resulting from fits and (b) transient position of the conduction band as function of pump-probe delay. (c) Cross correlation curve obtained from the conduction band at  0.57~eV above $E_{F}$ [\fig{raw_data}(d)]; the solid line corresponds to a double-exponential fit with two decay constants $\tau_{i}$ and $\tau_{ii}$; the dashed lines represent the two distinct contributions. (d)  $\tau_{ii}$ and $\tau_{i}$ as obtained from the fits  as function of energy  above $E_{F}$. The line represents a power law fit, which yields an exponent $(E-E_F)^{-1.5}$.
}
\label{hot_CB}
\end{figure}

The thermalization  and, eventually, the energy dissipation last over several picoseconds.
In order to investigate the dynamics of the hot electron gas quantitatively,  spectra and transients were fitted using Fermi-Dirac distributions and rate equations, respectively. The results are plotted in \fig{hot_CB}.  The electronic temperature $T_{el}$  rises  up to about 2000~K with a time constant of 260~fs. The maximum temperature and the time constant are in agreement with results of a previous x-ray diffraction study \cite {johnson2008} and estimates based on the fluence used here \cite{garl2010}, respectively.  The initial rise  is followed by a slow cooling of $T_{el}$ and energy dissipation from the electronic system to the lattice within about 5.9~ps. 

The cooling of the electronic system is accompanied by a corresponding shift of the band bottom on the same timescales, as shown in Fig. \ref{hot_CB}(b): Ignoring at present the first spike-like increase of the energy position, the average peak position shifts by 60 meV towards lower energies within a few hundreds of femtoseconds and subsequently relaxes back on a picosecond timescale. 
Similar observations were made very recently on Bi(111): following absorption of an infrared pump pulse both, surface and bulk states shifted to lower energy \cite{papalazarou2012}. The transient follows the electronic temperature like in our case, and the authors showed that the shift in energy is caused by an electronic effect. Since the origin could not be elucidated so far, we conjecture that the shift is caused by a charge redistribution due to the excitation of hot electrons.

The intensity cross correlation curves reveal two timescales $\tau_{i}$ and $\tau_{ii}$ depending on the electron energy, as can be seen in Figs.~\ref{hot_CB}(c) and (d): The long decay time $\tau_{ii}$ for electrons from the conduction band bottom or below reflect thermally excited states for delays larger than 1.5~ps and, thereby, the evolution of  $T_{el}$. The faster timescale  $\tau_{i}$ for energies within the conduction band (about 0.6~eV and higher) is caused by comparably fast scattering within the conduction band, which dominates the decay of these states.  The energy dependence corresponding to a power law with an exponent of $-1.5$ depends on details of the transient electronic distribution function \cite{rethfeld2002}. A detailed discussion is beyond the scope of this paper.

The dynamics found here for Bi(114) strongly resemble those observed very recently in the topolocial insulator Bi$_2$Se$_3$ \cite{sobota2012}: a fast intraband decay by electron-phonon scattering within the conduction band is followed by a much slower decay on a picosecond timescale for both the bulk conduction band and the metallic surface state. The second, long timescale is a consequence of the low scattering probability due  to the  low DOS at and above $E_F$ in Bi on one hand, and due to the energy difference between conduction band and surface state, which is large compared to typical phonon frequencies. 
In contrast to the above mentioned study, however, we find strong evidence for simultaneous emission of many  quanta of particular phonon modes from the observation of intensity and energy modulations of the photoelectron peaks, which can readily be recognized in \fig{raw_data}(c).

\begin{figure}[!h]
\centerline{\includegraphics[width = 0.44\textwidth]{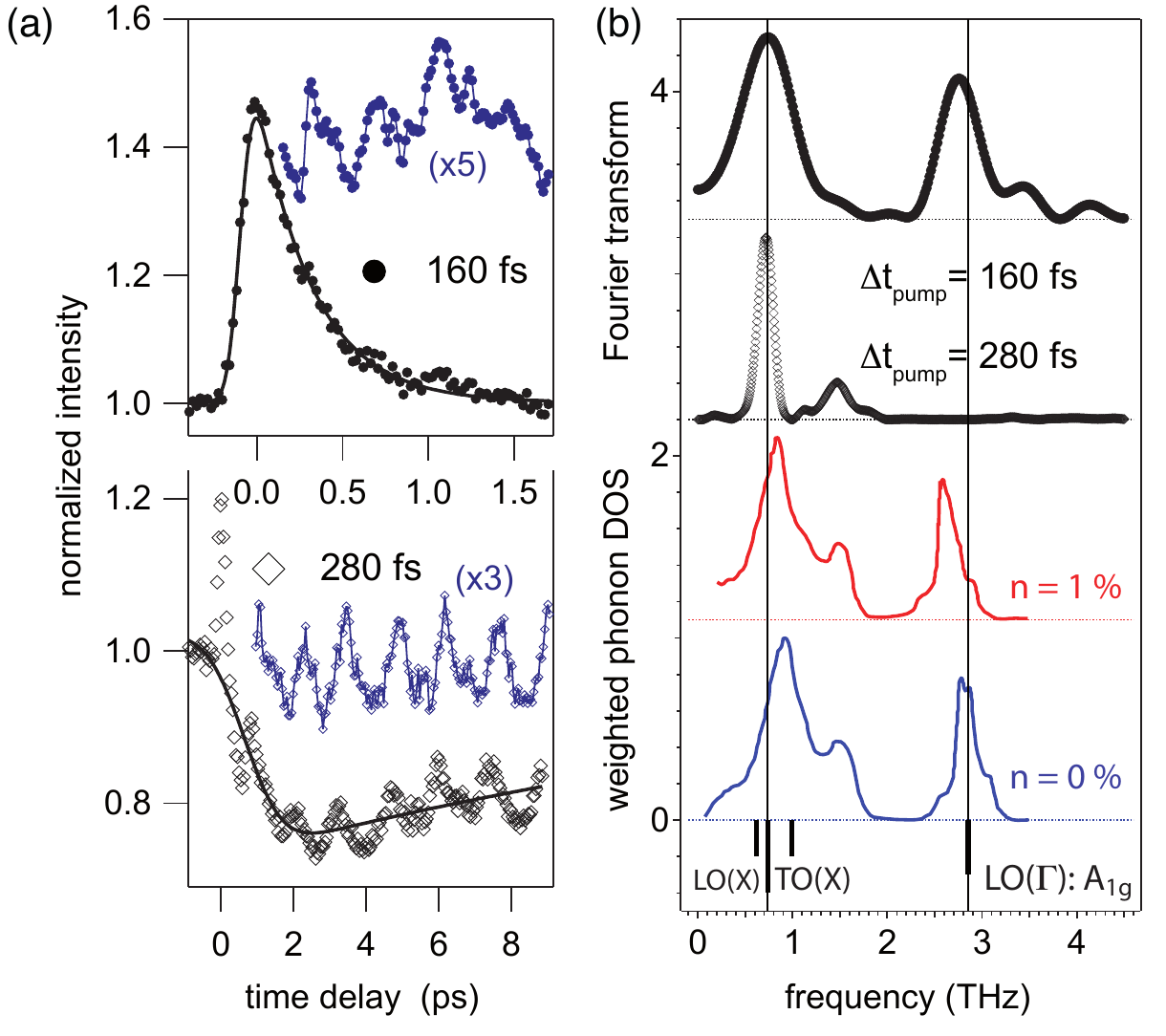}}
\caption{(Color online) Analysis of the modulations of photoemission intensities. (a) Transient photoemission intensity at the conduction band for a \emph{short} pump pulse (top panel) and at $E_F$ for a long pump pulse (bottom panel) as function of time delay. The solid lines denote rate equation fits. (b) As in (a) but after subtraction of the fitted rate equations: the oscillations of the intensity are clearly visible; the solid lines are cosine-functions at the dominant frequencies. (c) Fourier transform of the measured intensity cross correlations for the two different temporal pump pulse durations (solid symbols: $\Delta t_{pump}$~=~160~fs; open symbols: $\Delta t_{pump}$~=~280~fs). Bottom traces: calculated phonon DOS $F(\omega) d\omega$, taken from Ref.~\onlinecite{murray2007} and weighted by  $\omega^{-1}$ for the electronic ground state (blue) and for n~=~1$\%$ excited valence electrons (red). }
\label{oscillations_FT}
\end{figure}

In order to quantify these effects, the transient intensity changes were obtained by recording the intensity close to $E_F$ in case of the surface state and of the conduction band. The data were normalized to those at negative delay times. Two selected transients are displayed for two different pump pulse durations in \fig{oscillations_FT}(a). The cosine-like modulations, shown  in \fig{oscillations_FT}(b), are assigned to the excitations of coherent phonon modes. Such modes are well known to modulate the bulk and surface electronic states, allowing them to be directly observed in photoemission spectra \cite{perfetti2006, perfetti2008, Schmitt:2008p57, rettig2012}: the excitation of many electronic transitions and strong coupling to specific phonon modes drives phonons anharmonically, which results in a crystal structure closer to the simple cubic lattice in the present case. The atoms then start to perform damped oscillations around the new equilibrium positions. This phenomenon is called displacive excitation of coherent phonons if the initial excitation occurs within a time period shorter than half a phonon cycle.  As a consequence, the electronic spectral weight and energy position follow the changing atomic positions. Given that the electronic response is faster than the time period of the phonon mode, the phase of the coherent phonon is directly imprinted in the photoemission spectra. The amplitude of the oscillations in binding energy (\fig{hot_CB}b) is of the order of 20~meV, which is in agreement with photoemission data taken recently from Bi(111) under comparable conditions and corresponds to an atomic displacement of about 1-2~pm \cite{papalazarou2012}.

The Fourier transform (FT) of the measured oscillations, displayed as the top trace in \fig{oscillations_FT}(c) for a temporal pump pulse width $\Delta t_{pump}$ of 160~fs, reveals two dominant frequencies: The first frequency of 2.76(2)~THz is close to the softened longitudinal optical (LO) $A_{1g}$ mode, found at roughly  2.85~THz in numerous  time-resolved experiments \cite{cheng1990, hase1996,tinten03, fritz07, johnson2009} at the $\Gamma$-point of the bulk BZ. The second frequency of 0.72(1)~THz corresponds to an oscillation period of 1.39(1)~ps. The observation of this mode can be explained by the peculiar band structure of Bi(114) as will be discussed below.

Assuming the electron-phonon coupling strength to scale with inverse phonon frequency, the FT may be compared to $F(\omega) d\omega /\omega$ \cite{allen1987}, where $F(\omega) d\omega$ denotes the phonon DOS in the  frequency interval $\left[ \omega, \omega+d\omega \right]$. This function, which was calculated using the phonon DOS for two different electron excitation densities \cite{murray2007}, is displayed in \fig{oscillations_FT}(c). 
Two van Hove singularities dominate these spectra, one around 2.8 THz for the optical modes at $\Gamma$, and the second at about 0.7~THz corresponding to modes at the zone boundary along $\Gamma K X$ of the bulk BZ. 

We focus on the low-frequency mode at 0.72~THz. The excitation of this mode can either be the result of a strong decay channel of the high-frequency $A_{1g}$ mode or else result itself from a coherent electronic excitation.
In order to discriminate between both excitation pathways, the pump pulse duration was increased to $\Delta t =$~280~fs by introducing linear chirp by means of a grating compressor. Since half the temporal period of the $A_{1g}$ mode corresponds to about $T_{A1g}/2 \approx$~180~fs, the condition $\Delta t < T_{A1g}/2$ for coherent phonon excitation is no longer fulfilled. As a consequence, the $A_{1g}$ mode is suppressed [see \fig{oscillations_FT}(c)]. The low-frequency mode at 0.72~THz, on the other hand, persists giving evidence for a coherent excitation of this mode via electronic transitions. 

The fact that this mode can very efficiently be directly excited in Bi(114)  can be explained by  the low dimensionality of the surface \cite{gruener1994}. The strongly enhanced susceptibility of surface state electrons to excitations with momenta along the atomic chains increases the coupling significantly. Indeed, the strong phonon DOS around 0.7~THz in \fig{oscillations_FT}(c) is due to van Hove singularities in three  branches midway between two adjacent reciprocal lattice points along $\Gamma K X$. This corresponds to a standing wave in real space along the binary axis $[ 1\overline{1} 0]$, \textit{i.e.} along the atomic chains of the Bi(114)-surface. The character of these branches is acoustic in the long wavelength limit and changes to two transverse $TO(X)$ and one longitudinal optical $LO(X)$ branch at the border of the BZ\cite{murray2007}. Previously, a weak signal of about 1/6 of the $A_{1g}$ intensity at 0.68~THz was reported from time-resolved reflectivity measurements on Bi films \cite{wu2007} and interpreted as transverse-acoustic  mode at the X-point of the BZ in agreement with our interpretation of the Fourier spectra in the present work. Surprisingly and in contrast to a previous study on Bi(111) \cite{papalazarou2012}, the surface states are strongly modulated. This indicates that we are indeed dealing here with a strong phonon mode in the chains of Bi(114).

The wavelength of  4.54~\AA\  of these standing waves equals the atomic spacing in these chains, because the projection of the bulk $X$-point onto $k_x$ corresponds to the center of the second surface BZ as sketched in \fig{FSM}. 
This in turn means that the periodic lattice distortion must comprise at least the two topmost atomic rows of the chains (see \fig{FSM}), and that the atoms in the basis of the surface unit cell oscillate with respect to each other, which is equivalent to an optical phonon with infinite wavelength. So far no evidence was found  in our time-resolved experiments for  phonon modes introducing new spatial periodicities along the atomic rows.  The spin-orbit interaction, which lifts the spin degeneracy at the surface,  protects the metallic surface states against a phonon-driven charge-density wave transition, because efficient Fermi surface nesting is only allowed between states of the same spin helicity as shown convingly very recently for Pb wires on vicinal Si(557) \cite{tegenkamp2012}. 

In conclusion, the transient occupation of the bulk conduction band of Bi leads to the excitation of at least two coherent  optical phonon modes in the Bi(114) surface. Beside the $A_{1g}$ bulk phonon a second mode at 0.72~THz is found to strongly modulate  the surface electronic states on a picosecond time scale. This mode could be identified as an electronically induced coherent displacive excitation of a  standing wave along the atomic rows of the vicinal Bi(114) surface. 

We gratefully acknowledge J.I. Pascual and A. Strozecka for valuable discussions.
 This work was  supported  by  the Swiss National Science Foundation and  by  the Swiss National Science Foundation through the National Center of Competence in Research MUST.


\begin{thebibliography}{33}
\expandafter\ifx\csname natexlab\endcsname\relax\def\natexlab#1{#1}\fi
\expandafter\ifx\csname bibnamefont\endcsname\relax
  \def\bibnamefont#1{#1}\fi
\expandafter\ifx\csname bibfnamefont\endcsname\relax
  \def\bibfnamefont#1{#1}\fi
\expandafter\ifx\csname citenamefont\endcsname\relax
  \def\citenamefont#1{#1}\fi
\expandafter\ifx\csname url\endcsname\relax
  \def\url#1{\texttt{#1}}\fi
\expandafter\ifx\csname urlprefix\endcsname\relax\def\urlprefix{URL }\fi
\providecommand{\bibinfo}[2]{#2}
\providecommand{\eprint}[2][]{\url{#2}}

\bibitem[{\citenamefont{Hsieh et~al.}(2008)\citenamefont{Hsieh, Qian, Wray,
  Xia, Hor, Cava, and Hasan}}]{hsieh2008}
\bibinfo{author}{\bibfnamefont{D.}~\bibnamefont{Hsieh}},
  \bibinfo{author}{\bibfnamefont{D.}~\bibnamefont{Qian}},
  \bibinfo{author}{\bibfnamefont{L.}~\bibnamefont{Wray}},
  \bibinfo{author}{\bibfnamefont{Y.}~\bibnamefont{Xia}},
  \bibinfo{author}{\bibfnamefont{Y.~S.} \bibnamefont{Hor}},
  \bibinfo{author}{\bibfnamefont{R.~J.} \bibnamefont{Cava}}, \bibnamefont{and}
  \bibinfo{author}{\bibfnamefont{M.~Z.} \bibnamefont{Hasan}},
  \bibinfo{journal}{Nature} \textbf{\bibinfo{volume}{452}},
  \bibinfo{pages}{970} (\bibinfo{year}{2008}).

\bibitem[{\citenamefont{Hasan and Kane}(2010)}]{hasan2010}
\bibinfo{author}{\bibfnamefont{M.~Z.} \bibnamefont{Hasan}} \bibnamefont{and}
  \bibinfo{author}{\bibfnamefont{C.~L.} \bibnamefont{Kane}},
  \bibinfo{journal}{Rev. Mod. Phys.} \textbf{\bibinfo{volume}{82}},
  \bibinfo{pages}{3045} (\bibinfo{year}{2010}).

\bibitem[{\citenamefont{Hengsberger et~al.}(2000)\citenamefont{Hengsberger,
  Segovia, Garnier, Purdie, and Baer}}]{hengsberger2000JEPB}
\bibinfo{author}{\bibfnamefont{M.}~\bibnamefont{Hengsberger}},
  \bibinfo{author}{\bibfnamefont{P.}~\bibnamefont{Segovia}},
  \bibinfo{author}{\bibfnamefont{M.}~\bibnamefont{Garnier}},
  \bibinfo{author}{\bibfnamefont{D.}~\bibnamefont{Purdie}}, \bibnamefont{and}
  \bibinfo{author}{\bibfnamefont{Y.}~\bibnamefont{Baer}},
  \bibinfo{journal}{Eur. Phys. J. B} \textbf{\bibinfo{volume}{17}},
  \bibinfo{pages}{603} (\bibinfo{year}{2000}).

\bibitem[{\citenamefont{Agergaard et~al.}(2001)\citenamefont{Agergaard,
  Sondergaard, Li, Nielsen, Hoffmann, Li, and {P}h. Hofmann}}]{agergaard2001}
\bibinfo{author}{\bibfnamefont{S.}~\bibnamefont{Agergaard}},
  \bibinfo{author}{\bibfnamefont{C.}~\bibnamefont{Sondergaard}},
  \bibinfo{author}{\bibfnamefont{H.}~\bibnamefont{Li}},
  \bibinfo{author}{\bibfnamefont{M.~B.} \bibnamefont{Nielsen}},
  \bibinfo{author}{\bibfnamefont{S.~V.} \bibnamefont{Hoffmann}},
  \bibinfo{author}{\bibfnamefont{Z.}~\bibnamefont{Li}}, \bibnamefont{and}
  \bibinfo{author}{\bibnamefont{{P}h. Hofmann}}, \bibinfo{journal}{New J. of
  Phys.} \textbf{\bibinfo{volume}{3}}, \bibinfo{pages}{15.1}
  (\bibinfo{year}{2001}).

\bibitem[{\citenamefont{Ast and H\"ochst}(2001)}]{ast2001}
\bibinfo{author}{\bibfnamefont{C.~R.} \bibnamefont{Ast}} \bibnamefont{and}
  \bibinfo{author}{\bibfnamefont{H.}~\bibnamefont{H\"ochst}},
  \bibinfo{journal}{Phys. Rev. Lett.} \textbf{\bibinfo{volume}{87}},
  \bibinfo{pages}{177602} (\bibinfo{year}{2001}).

\bibitem[{\citenamefont{{P}h. Hofmann}(2006)}]{hofmann2006}
\bibinfo{author}{\bibnamefont{{P}h. Hofmann}}, \bibinfo{journal}{Progress in
  Surface Science 81} \textbf{\bibinfo{volume}{81}}, \bibinfo{pages}{191}
  (\bibinfo{year}{2006}).

\bibitem[{\citenamefont{Wells et~al.}(2009)\citenamefont{Wells, Dil, Meier,
  Lobo-Checa, Petrov, Osterwalder, Ugeda, Fernandez-Torrente, Pascual, Rienks
  et~al.}}]{wells09}
\bibinfo{author}{\bibfnamefont{J.~W.} \bibnamefont{Wells}},
  \bibinfo{author}{\bibfnamefont{J.~H.} \bibnamefont{Dil}},
  \bibinfo{author}{\bibfnamefont{F.}~\bibnamefont{Meier}},
  \bibinfo{author}{\bibfnamefont{J.}~\bibnamefont{Lobo-Checa}},
  \bibinfo{author}{\bibfnamefont{V.~N.} \bibnamefont{Petrov}},
  \bibinfo{author}{\bibfnamefont{J.}~\bibnamefont{Osterwalder}},
  \bibinfo{author}{\bibfnamefont{M.~M.} \bibnamefont{Ugeda}},
  \bibinfo{author}{\bibfnamefont{I.}~\bibnamefont{Fernandez-Torrente}},
  \bibinfo{author}{\bibfnamefont{J.~I.} \bibnamefont{Pascual}},
  \bibinfo{author}{\bibfnamefont{E.~D.~L.} \bibnamefont{Rienks}},
  \bibnamefont{et~al.}, \bibinfo{journal}{Phys. Rev. Lett.}
  \textbf{\bibinfo{volume}{102}}, \bibinfo{pages}{096802}
  (\bibinfo{year}{2009}).

\bibitem[{\citenamefont{Gr\"{u}ner}(1994)}]{gruener1994}
\bibinfo{author}{\bibfnamefont{G.}~\bibnamefont{Gr\"{u}ner}},
  \emph{\bibinfo{title}{Density Waves in Solids}}, vol.~\bibinfo{volume}{89} of
  \emph{\bibinfo{series}{Frontiers in Physics}} (\bibinfo{publisher}{Addison
  Wesley Publishing, Reading, Menlo Park, New York}, \bibinfo{year}{1994}).

\bibitem[{\citenamefont{Ast and H\"ochst}(2003)}]{ast2003}
\bibinfo{author}{\bibfnamefont{C.~R.} \bibnamefont{Ast}} \bibnamefont{and}
  \bibinfo{author}{\bibfnamefont{H.}~\bibnamefont{H\"ochst}},
  \bibinfo{journal}{Phys. Rev. Lett.} \textbf{\bibinfo{volume}{90}},
  \bibinfo{pages}{016403} (\bibinfo{year}{2003}).

\bibitem[{\citenamefont{Kim et~al.}(2005)\citenamefont{Kim, Wells, Kirkegaard,
  Li, Hoffmann, Gayone, Fernandez-Torrente, H\"aberle, Pascual, Moore
  et~al.}}]{kim2005}
\bibinfo{author}{\bibfnamefont{T.~K.} \bibnamefont{Kim}},
  \bibinfo{author}{\bibfnamefont{J.}~\bibnamefont{Wells}},
  \bibinfo{author}{\bibfnamefont{C.}~\bibnamefont{Kirkegaard}},
  \bibinfo{author}{\bibfnamefont{Z.}~\bibnamefont{Li}},
  \bibinfo{author}{\bibfnamefont{S.~V.} \bibnamefont{Hoffmann}},
  \bibinfo{author}{\bibfnamefont{J.~E.} \bibnamefont{Gayone}},
  \bibinfo{author}{\bibfnamefont{I.}~\bibnamefont{Fernandez-Torrente}},
  \bibinfo{author}{\bibfnamefont{P.}~\bibnamefont{H\"aberle}},
  \bibinfo{author}{\bibfnamefont{J.~I.} \bibnamefont{Pascual}},
  \bibinfo{author}{\bibfnamefont{K.~T.} \bibnamefont{Moore}},
  \bibnamefont{et~al.}, \bibinfo{journal}{Phys. Rev. B}
  \textbf{\bibinfo{volume}{72}}, \bibinfo{pages}{085440}
  (\bibinfo{year}{2005}).

\bibitem[{\citenamefont{Cheng et~al.}(1990)\citenamefont{Cheng, Brorson,
  Kazeroonian, Moodera, Dresselhaus, Dresselhaus, and Ippen}}]{cheng1990}
\bibinfo{author}{\bibfnamefont{T.~K.} \bibnamefont{Cheng}},
  \bibinfo{author}{\bibfnamefont{S.~D.} \bibnamefont{Brorson}},
  \bibinfo{author}{\bibfnamefont{A.~S.} \bibnamefont{Kazeroonian}},
  \bibinfo{author}{\bibfnamefont{J.~S.} \bibnamefont{Moodera}},
  \bibinfo{author}{\bibfnamefont{G.}~\bibnamefont{Dresselhaus}},
  \bibinfo{author}{\bibfnamefont{M.~S.} \bibnamefont{Dresselhaus}},
  \bibnamefont{and} \bibinfo{author}{\bibfnamefont{E.~P.} \bibnamefont{Ippen}},
  \bibinfo{journal}{Appl. Phys. Lett.} \textbf{\bibinfo{volume}{57}},
  \bibinfo{pages}{1004} (\bibinfo{year}{1990}).

\bibitem[{\citenamefont{Hase et~al.}(1996)\citenamefont{Hase, Mizoguchi,
  Harima, Nakashima, Tani, Sakai, and Hangyo}}]{hase1996}
\bibinfo{author}{\bibfnamefont{M.}~\bibnamefont{Hase}},
  \bibinfo{author}{\bibfnamefont{K.}~\bibnamefont{Mizoguchi}},
  \bibinfo{author}{\bibfnamefont{H.}~\bibnamefont{Harima}},
  \bibinfo{author}{\bibfnamefont{S.}~\bibnamefont{Nakashima}},
  \bibinfo{author}{\bibfnamefont{M.}~\bibnamefont{Tani}},
  \bibinfo{author}{\bibfnamefont{K.}~\bibnamefont{Sakai}}, \bibnamefont{and}
  \bibinfo{author}{\bibfnamefont{M.}~\bibnamefont{Hangyo}},
  \bibinfo{journal}{Appl. Phys. Lett.} \textbf{\bibinfo{volume}{100}},
  \bibinfo{pages}{2474} (\bibinfo{year}{1996}).

\bibitem[{\citenamefont{Sokolowski-Tinten
  et~al.}(2003)\citenamefont{Sokolowski-Tinten, Blome, Blums, Cavalleri,
  Dietrich, Tarasevitch, Uschmann, F\"orster, Kammler, von Hoegen
  et~al.}}]{tinten03}
\bibinfo{author}{\bibfnamefont{K.}~\bibnamefont{Sokolowski-Tinten}},
  \bibinfo{author}{\bibfnamefont{C.}~\bibnamefont{Blome}},
  \bibinfo{author}{\bibfnamefont{J.}~\bibnamefont{Blums}},
  \bibinfo{author}{\bibfnamefont{A.}~\bibnamefont{Cavalleri}},
  \bibinfo{author}{\bibfnamefont{C.}~\bibnamefont{Dietrich}},
  \bibinfo{author}{\bibfnamefont{A.}~\bibnamefont{Tarasevitch}},
  \bibinfo{author}{\bibfnamefont{I.}~\bibnamefont{Uschmann}},
  \bibinfo{author}{\bibfnamefont{E.}~\bibnamefont{F\"orster}},
  \bibinfo{author}{\bibfnamefont{M.}~\bibnamefont{Kammler}},
  \bibinfo{author}{\bibfnamefont{M.~H.} \bibnamefont{von Hoegen}},
  \bibnamefont{et~al.}, \bibinfo{journal}{Nature}
  \textbf{\bibinfo{volume}{422}}, \bibinfo{pages}{287} (\bibinfo{year}{2003}).

\bibitem[{\citenamefont{Fritz et~al.}(2007)\citenamefont{Fritz, Reis, Adams,
  Akre, Arthur, Blome, Bucksbaum, Cavalieri, Engemann, Fahy et~al.}}]{fritz07}
\bibinfo{author}{\bibfnamefont{D.~M.} \bibnamefont{Fritz}},
  \bibinfo{author}{\bibfnamefont{D.~A.} \bibnamefont{Reis}},
  \bibinfo{author}{\bibfnamefont{B.}~\bibnamefont{Adams}},
  \bibinfo{author}{\bibfnamefont{R.~A.} \bibnamefont{Akre}},
  \bibinfo{author}{\bibfnamefont{J.}~\bibnamefont{Arthur}},
  \bibinfo{author}{\bibfnamefont{C.}~\bibnamefont{Blome}},
  \bibinfo{author}{\bibfnamefont{P.~H.} \bibnamefont{Bucksbaum}},
  \bibinfo{author}{\bibfnamefont{A.~L.} \bibnamefont{Cavalieri}},
  \bibinfo{author}{\bibfnamefont{S.}~\bibnamefont{Engemann}},
  \bibinfo{author}{\bibfnamefont{S.}~\bibnamefont{Fahy}}, \bibnamefont{et~al.},
  \bibinfo{journal}{Science} \textbf{\bibinfo{volume}{315}},
  \bibinfo{pages}{633} (\bibinfo{year}{2007}).

\bibitem[{\citenamefont{Johnson et~al.}(2009)\citenamefont{Johnson, Beaud,
  Vorobeva, Milne, Murray, Fahy, and Ingold}}]{johnson2009}
\bibinfo{author}{\bibfnamefont{S.~L.} \bibnamefont{Johnson}},
  \bibinfo{author}{\bibfnamefont{P.}~\bibnamefont{Beaud}},
  \bibinfo{author}{\bibfnamefont{E.}~\bibnamefont{Vorobeva}},
  \bibinfo{author}{\bibfnamefont{C.~J.} \bibnamefont{Milne}},
  \bibinfo{author}{\bibfnamefont{E.~D.} \bibnamefont{Murray}},
  \bibinfo{author}{\bibfnamefont{S.}~\bibnamefont{Fahy}}, \bibnamefont{and}
  \bibinfo{author}{\bibfnamefont{G.}~\bibnamefont{Ingold}},
  \bibinfo{journal}{Phys. Rev. Lett.} \textbf{\bibinfo{volume}{102}},
  \bibinfo{pages}{175503} (\bibinfo{year}{2009}).

\bibitem[{\citenamefont{Leuenberger et~al.}(2011)\citenamefont{Leuenberger,
  Yanagisawa, Roth, Osterwalder, and Hengsberger}}]{leuenberger2011}
\bibinfo{author}{\bibfnamefont{D.}~\bibnamefont{Leuenberger}},
  \bibinfo{author}{\bibfnamefont{H.}~\bibnamefont{Yanagisawa}},
  \bibinfo{author}{\bibfnamefont{S.}~\bibnamefont{Roth}},
  \bibinfo{author}{\bibfnamefont{J.}~\bibnamefont{Osterwalder}},
  \bibnamefont{and}
  \bibinfo{author}{\bibfnamefont{M.}~\bibnamefont{Hengsberger}},
  \bibinfo{journal}{Phys. Rev. B} \textbf{\bibinfo{volume}{84}},
  \bibinfo{pages}{125107} (\bibinfo{year}{2011}).

\bibitem[{\citenamefont{Greber et~al.}(1997)\citenamefont{Greber, Raetzo,
  Kreutz, Schwaller, Deichmann, Wetli, and Osterwalder}}]{Greber:1997p68}
\bibinfo{author}{\bibfnamefont{T.}~\bibnamefont{Greber}},
  \bibinfo{author}{\bibfnamefont{O.}~\bibnamefont{Raetzo}},
  \bibinfo{author}{\bibfnamefont{T.}~\bibnamefont{Kreutz}},
  \bibinfo{author}{\bibfnamefont{P.}~\bibnamefont{Schwaller}},
  \bibinfo{author}{\bibfnamefont{W.}~\bibnamefont{Deichmann}},
  \bibinfo{author}{\bibfnamefont{E.}~\bibnamefont{Wetli}}, \bibnamefont{and}
  \bibinfo{author}{\bibfnamefont{J.}~\bibnamefont{Osterwalder}},
  \bibinfo{journal}{Rev. Sci. Instrum.} \textbf{\bibinfo{volume}{68}},
  \bibinfo{pages}{4549} (\bibinfo{year}{1997}).

\bibitem[{\citenamefont{Hengsberger et~al.}(2008)\citenamefont{Hengsberger,
  Baumberger, Neff, Greber, and Osterwalder}}]{hengsberger08}
\bibinfo{author}{\bibfnamefont{M.}~\bibnamefont{Hengsberger}},
  \bibinfo{author}{\bibfnamefont{F.}~\bibnamefont{Baumberger}},
  \bibinfo{author}{\bibfnamefont{H.~J.} \bibnamefont{Neff}},
  \bibinfo{author}{\bibfnamefont{T.}~\bibnamefont{Greber}}, \bibnamefont{and}
  \bibinfo{author}{\bibfnamefont{J.}~\bibnamefont{Osterwalder}},
  \bibinfo{journal}{Phys. Rev. B} \textbf{\bibinfo{volume}{77}},
  \bibinfo{pages}{085425} (\bibinfo{year}{2008}).

\bibitem[{\citenamefont{Liu and Allen}(1995)}]{liu1995}
\bibinfo{author}{\bibfnamefont{Y.}~\bibnamefont{Liu}} \bibnamefont{and}
  \bibinfo{author}{\bibfnamefont{R.~E.} \bibnamefont{Allen}},
  \bibinfo{journal}{Phys. Rev. B} \textbf{\bibinfo{volume}{52}},
  \bibinfo{pages}{1566} (\bibinfo{year}{1995}).

\bibitem[{foo()}]{footnoteCBM}
\bibinfo{note}{Note that the global bulk conduction band minimum is below the
  Fermi energy. We refer to the band bottom here as being the energy minimum of
  the bulk conduction band along [114], as observed in the photoemission
  spectra for this geometry.}

\bibitem[{\citenamefont{Johnson et~al.}(2008)\citenamefont{Johnson, Beaud,
  Milne, Krasniqi, Zijlstra, Gracia, Kaiser, Grolimund, Abela, and
  Ingold}}]{johnson2008}
\bibinfo{author}{\bibfnamefont{S.~L.} \bibnamefont{Johnson}},
  \bibinfo{author}{\bibfnamefont{P.}~\bibnamefont{Beaud}},
  \bibinfo{author}{\bibfnamefont{C.~J.} \bibnamefont{Milne}},
  \bibinfo{author}{\bibfnamefont{F.~S.} \bibnamefont{Krasniqi}},
  \bibinfo{author}{\bibfnamefont{E.~S.} \bibnamefont{Zijlstra}},
  \bibinfo{author}{\bibfnamefont{M.~E.} \bibnamefont{Garcia}},
  \bibinfo{author}{\bibfnamefont{M.}~\bibnamefont{Kaiser}},
  \bibinfo{author}{\bibfnamefont{D.}~\bibnamefont{Grolimund}},
  \bibinfo{author}{\bibfnamefont{R.}~\bibnamefont{Abela}}, \bibnamefont{and}
  \bibinfo{author}{\bibfnamefont{G.}~\bibnamefont{Ingold}},
  \bibinfo{journal}{Phys. Rev. Lett.} \textbf{\bibinfo{volume}{100}},
  \bibinfo{pages}{155501} (\bibinfo{year}{2008}).

\bibitem[{\citenamefont{Boschetto et~al.}(2010)\citenamefont{Boschetto, Garl,
  and Rousse}}]{garl2010}
\bibinfo{author}{\bibfnamefont{D.}~\bibnamefont{Boschetto}},
  \bibinfo{author}{\bibfnamefont{T.}~\bibnamefont{Garl}}, \bibnamefont{and}
  \bibinfo{author}{\bibfnamefont{A.}~\bibnamefont{Rousse}},
  \bibinfo{journal}{Journal of Modern Optics} \textbf{\bibinfo{volume}{57}}
  (\bibinfo{year}{2010}).

\bibitem[{\citenamefont{Papalazarou et~al.}(2012)\citenamefont{Papalazarou,
  Faure, Mauchain, Marsi, Taleb-Ibrahimi, Reshetnyak, van Roekeghem, Timrov,
  Vast, Arnaud et~al.}}]{papalazarou2012}
\bibinfo{author}{\bibfnamefont{E.}~\bibnamefont{Papalazarou}},
  \bibinfo{author}{\bibfnamefont{J.}~\bibnamefont{Faure}},
  \bibinfo{author}{\bibfnamefont{J.}~\bibnamefont{Mauchain}},
  \bibinfo{author}{\bibfnamefont{M.}~\bibnamefont{Marsi}},
  \bibinfo{author}{\bibfnamefont{A.}~\bibnamefont{Taleb-Ibrahimi}},
  \bibinfo{author}{\bibfnamefont{I.}~\bibnamefont{Reshetnyak}},
  \bibinfo{author}{\bibfnamefont{A.}~\bibnamefont{van Roekeghem}},
  \bibinfo{author}{\bibfnamefont{I.}~\bibnamefont{Timrov}},
  \bibinfo{author}{\bibfnamefont{N.}~\bibnamefont{Vast}},
  \bibinfo{author}{\bibfnamefont{B.}~\bibnamefont{Arnaud}},
  \bibnamefont{et~al.}, \bibinfo{journal}{Phys. Rev. Lett.}
  \textbf{\bibinfo{volume}{108}}, \bibinfo{pages}{256808}
  (\bibinfo{year}{2012}).

\bibitem[{\citenamefont{Rethfeld et~al.}(2002)\citenamefont{Rethfeld, Kaiser,
  Vicanek, and Simon}}]{rethfeld2002}
\bibinfo{author}{\bibfnamefont{B.}~\bibnamefont{Rethfeld}},
  \bibinfo{author}{\bibfnamefont{A.}~\bibnamefont{Kaiser}},
  \bibinfo{author}{\bibfnamefont{M.}~\bibnamefont{Vicanek}}, \bibnamefont{and}
  \bibinfo{author}{\bibfnamefont{G.}~\bibnamefont{Simon}},
  \bibinfo{journal}{Phys. Rev. B} \textbf{\bibinfo{volume}{65}},
  \bibinfo{pages}{214303} (\bibinfo{year}{2002}).

\bibitem[{\citenamefont{Sobota et~al.}(2012)\citenamefont{Sobota, Yang,
  Analytis, Chen, Fisher, Kirchmann, and Shen}}]{sobota2012}
\bibinfo{author}{\bibfnamefont{J.~A.} \bibnamefont{Sobota}},
  \bibinfo{author}{\bibfnamefont{S.}~\bibnamefont{Yang}},
  \bibinfo{author}{\bibfnamefont{J.~G.} \bibnamefont{Analytis}},
  \bibinfo{author}{\bibfnamefont{Y.~L.} \bibnamefont{Chen}},
  \bibinfo{author}{\bibfnamefont{I.~R.} \bibnamefont{Fisher}},
  \bibinfo{author}{\bibfnamefont{P.~S.} \bibnamefont{Kirchmann}},
  \bibnamefont{and} \bibinfo{author}{\bibfnamefont{Z.~X.} \bibnamefont{Shen}},
  \bibinfo{journal}{Phys. Rev. Lett.} \textbf{\bibinfo{volume}{108}},
  \bibinfo{pages}{117403} (\bibinfo{year}{2012}).

\bibitem[{\citenamefont{Perfetti et~al.}(2006)\citenamefont{Perfetti, Loukakos,
  Lisowski, Bovensiepen, Berger, Biermann, Cornaglia, Georges, and
  Wolf}}]{perfetti2006}
\bibinfo{author}{\bibfnamefont{L.}~\bibnamefont{Perfetti}},
  \bibinfo{author}{\bibfnamefont{P.~A.} \bibnamefont{Loukakos}},
  \bibinfo{author}{\bibfnamefont{M.}~\bibnamefont{Lisowski}},
  \bibinfo{author}{\bibfnamefont{U.}~\bibnamefont{Bovensiepen}},
  \bibinfo{author}{\bibfnamefont{H.}~\bibnamefont{Berger}},
  \bibinfo{author}{\bibfnamefont{S.}~\bibnamefont{Biermann}},
  \bibinfo{author}{\bibfnamefont{P.~S.} \bibnamefont{Cornaglia}},
  \bibinfo{author}{\bibfnamefont{A.}~\bibnamefont{Georges}}, \bibnamefont{and}
  \bibinfo{author}{\bibfnamefont{M.}~\bibnamefont{Wolf}},
  \bibinfo{journal}{Phys. Rev. Lett.} \textbf{\bibinfo{volume}{97}},
  \bibinfo{pages}{067402} (\bibinfo{year}{2006}).

\bibitem[{\citenamefont{Perfetti et~al.}(2008)\citenamefont{Perfetti, Loukakos,
  Lisowski, Bovensiepen, Wolf, Berger, Biermann, and Georges}}]{perfetti2008}
\bibinfo{author}{\bibfnamefont{L.}~\bibnamefont{Perfetti}},
  \bibinfo{author}{\bibfnamefont{P.~A.} \bibnamefont{Loukakos}},
  \bibinfo{author}{\bibfnamefont{M.}~\bibnamefont{Lisowski}},
  \bibinfo{author}{\bibfnamefont{U.}~\bibnamefont{Bovensiepen}},
  \bibinfo{author}{\bibfnamefont{M.}~\bibnamefont{Wolf}},
  \bibinfo{author}{\bibfnamefont{H.}~\bibnamefont{Berger}},
  \bibinfo{author}{\bibfnamefont{S.}~\bibnamefont{Biermann}}, \bibnamefont{and}
  \bibinfo{author}{\bibfnamefont{A.}~\bibnamefont{Georges}},
  \bibinfo{journal}{New J. Phys.} \textbf{\bibinfo{volume}{10}},
  \bibinfo{pages}{053019} (\bibinfo{year}{2008}).

\bibitem[{\citenamefont{Schmitt et~al.}(2008)\citenamefont{Schmitt, Kirchmann,
  Bovensiepen, Moore, Rettig, Krenz, Chu, Ru, Perfetti, Lu
  et~al.}}]{Schmitt:2008p57}
\bibinfo{author}{\bibfnamefont{F.}~\bibnamefont{Schmitt}},
  \bibinfo{author}{\bibfnamefont{P.~S.} \bibnamefont{Kirchmann}},
  \bibinfo{author}{\bibfnamefont{U.}~\bibnamefont{Bovensiepen}},
  \bibinfo{author}{\bibfnamefont{R.~G.} \bibnamefont{Moore}},
  \bibinfo{author}{\bibfnamefont{L.}~\bibnamefont{Rettig}},
  \bibinfo{author}{\bibfnamefont{M.}~\bibnamefont{Krenz}},
  \bibinfo{author}{\bibfnamefont{J.-H.} \bibnamefont{Chu}},
  \bibinfo{author}{\bibfnamefont{N.}~\bibnamefont{Ru}},
  \bibinfo{author}{\bibfnamefont{L.}~\bibnamefont{Perfetti}},
  \bibinfo{author}{\bibfnamefont{D.~H.} \bibnamefont{Lu}},
  \bibnamefont{et~al.}, \bibinfo{journal}{Science}
  \textbf{\bibinfo{volume}{321}}, \bibinfo{pages}{1649} (\bibinfo{year}{2008}).

\bibitem[{\citenamefont{Rettig et~al.}(2012)\citenamefont{Rettig, Kirchmann,
  and Bovensiepen}}]{rettig2012}
\bibinfo{author}{\bibfnamefont{L.}~\bibnamefont{Rettig}},
  \bibinfo{author}{\bibfnamefont{P.~S.} \bibnamefont{Kirchmann}},
  \bibnamefont{and}
  \bibinfo{author}{\bibfnamefont{U.}~\bibnamefont{Bovensiepen}},
  \bibinfo{journal}{New J. Phys.} \textbf{\bibinfo{volume}{14}},
  \bibinfo{pages}{023047} (\bibinfo{year}{2012}).

\bibitem[{\citenamefont{Murray et~al.}(2007)\citenamefont{Murray, Fahy,
  Prendergast, Ogitsu, Fritz, and Reis}}]{murray2007}
\bibinfo{author}{\bibfnamefont{E.~D.} \bibnamefont{Murray}},
  \bibinfo{author}{\bibfnamefont{S.}~\bibnamefont{Fahy}},
  \bibinfo{author}{\bibfnamefont{D.}~\bibnamefont{Prendergast}},
  \bibinfo{author}{\bibfnamefont{T.}~\bibnamefont{Ogitsu}},
  \bibinfo{author}{\bibfnamefont{D.~M.} \bibnamefont{Fritz}}, \bibnamefont{and}
  \bibinfo{author}{\bibfnamefont{D.~A.} \bibnamefont{Reis}},
  \bibinfo{journal}{Phys. Rev. B} \textbf{\bibinfo{volume}{75}},
  \bibinfo{pages}{184301} (\bibinfo{year}{2007}).

\bibitem[{\citenamefont{Allen}(1987)}]{allen1987}
\bibinfo{author}{\bibfnamefont{P.~B.} \bibnamefont{Allen}},
  \bibinfo{journal}{Phys. Rev. Lett.} \textbf{\bibinfo{volume}{59}},
  \bibinfo{pages}{1460} (\bibinfo{year}{1987}).

\bibitem[{\citenamefont{Wu and Xu}(2007)}]{wu2007}
\bibinfo{author}{\bibfnamefont{A.}~\bibnamefont{Wu}} \bibnamefont{and}
  \bibinfo{author}{\bibfnamefont{X.}~\bibnamefont{Xu}}, \bibinfo{journal}{Appl.
  Surf. Sci.} \textbf{\bibinfo{volume}{253}}, \bibinfo{pages}{6301}
  (\bibinfo{year}{2007}).

\bibitem[{\citenamefont{Tegenkamp et~al.}(2012)\citenamefont{Tegenkamp,
  L\"ukermann, Pfn\"ur, Slomski, Landolt, and Dil}}]{tegenkamp2012}
\bibinfo{author}{\bibfnamefont{C.}~\bibnamefont{Tegenkamp}},
  \bibinfo{author}{\bibfnamefont{D.}~\bibnamefont{L\"ukermann}},
  \bibinfo{author}{\bibfnamefont{H.}~\bibnamefont{Pfn\"ur}},
  \bibinfo{author}{\bibfnamefont{B.}~\bibnamefont{Slomski}},
  \bibinfo{author}{\bibfnamefont{G.}~\bibnamefont{Landolt}}, \bibnamefont{and}
  \bibinfo{author}{\bibfnamefont{J.~H.} \bibnamefont{Dil}},
  \bibinfo{journal}{Phys. Rev. Lett.} \textbf{\bibinfo{volume}{109}},
  \bibinfo{pages}{266401} (\bibinfo{year}{2012}).

\end{thebibliography}
\end{document}